\documentstyle[graphicx,proceedings,numreferences]{balkan}
\newcommand{\ga}{\alpha}
\newcommand{\gb}{\beta}
\newcommand{\gc}{\gamma}
\newcommand{\gd}{\delta}

\newcommand{\gl}{\lambda}

\begin{opening}
 \title{THE IN-MEDIUM FEW-BODY PROBLEM}
 \author{ S. A. Sofianos$^1$}
 \author{ M. Beyer$^{1,2}$}
\institute{
{\it $^1$Physics Department, University of South Africa,\\
 P.O.  Box 392, Pretoria 0003, South Africa\\
 {\it $^2$Fachbereich Physik, Universit\"at Rostock,\\
     18051 Rostock, Germany}
}}

\date{\today}
\end{opening}
\begin{document}
\begin{abstract}
We are concerned with few-particle correlations in a fermionic
  system at finite temperature and density.  Within the many-body
  Green functions formalism the description of correlations is
  provided by the Dyson equation approach that leads to effective
  few-body equations. They contain the dominant medium effects, which
  are self energy corrections and the Pauli blocking.  Hence the
  effective two-body interactions between quasiparticles are
  momentum/energy-dependent and therefore they can be
  usesed in the medium modified, momentum space, 
  integral AGS equations for three- and four-body  systems.
 To investigate correlations and clusters
  beyond four-body, we employ, instead, the configuration 
  space two-variable integro-differential  equations (IDEA) 
  for $A$-body bound systems which are based on Hyperspherical
  Harmonics and the Faddeev decomposition of the wave function
  in two-body amplitudes.  This requires the transformation 
  of the energy dependent two-body interactions to equivalent local,
  energy independent, ones. To achieve this we  use
  inverse scattering  techniques the resulting interactions 
  being, on-- and (to all practical purposes) off--shell
  equivalent to the energy dependent potentials. In this way 
  we obtain binding energy results  for the 2--, 3--, 4--,  
  and 16--particle in a   medium at a finite temperature 
  and various densities. Several aspects of the problem are 
  discussed and the behavior of the potential surfaces obtained 
  in the extreme adiabatic approximation, below and above the 
  Mott  transition, is investigated.
\end{abstract}

\section{Introduction}

The properties of nuclear matter can be studied using various methods
such as path integral methods, computer simulations, or perturbation
theory using many-body Green functions \cite{fet71}.  In the latter,
which will be followed here, the equations of state may be expressed
in terms of many-body Green functions and the inclusion of
correlations is provided by the Dyson equation approach developed for
this  purpose in~\cite{sk71,sk73,sk90,sk91,sk94,sk96}, and reviewed
in~\cite{duk98}.  Within the Dyson equation approach the familiar
in-medium two-body equations, known as Galitskii-Feynman or
Bethe-Goldstone equations (depending on details), can be
derived~\cite{fet71}. Recently, these equations have been extended to
investigate correlations and clusters of more than two particles
and applied to various aspects of fragmentation in a heavy ion
collision at intermediate
energies~\cite{Beyer:1996rx,bey97,Beyer:1997sf,Beyer:1999tm,Beyer:1999zx,Kuhrts:2000jz,Beyer:1999xv,Beyer:2000ds,Kuhrts:2001zs,Beyer:2003zz}.
For three- and four-body correlations this approach leads to modified
momentum space integral equations of the Alt-Grassberger-Sandhas
(AGS)~\cite{ags1,ags2} type.

For correlations with more particles ($A>4$), relevant e.g. in
fragmentation of heavy ion collisions at intermediate
energies~\cite{Hudan:2002tn}, the momentum space integral equations are
not appropriate and thus one has to resort to effective two-body
interactions for the clusters involved and consider the system under
consideration as a few-body system. For instance, the $^{16}$O may be
considered as a 4-$\alpha$ system interacting via the $\alpha-\alpha$
interaction.  In such a formalism, however, one faces the problem of
how to construct the inter-cluster interactions which is a difficult
task to implement. Apart from this, most of the existing and well
established theories for the description of the medium-light nuclei,
$5\le A\le 20$ say, are in configuration space and are based on
energy-independent two-body potentials.  Therefore, the energy
dependent effective interactions can not be directly employed and thus
one should resort to some localization procedure which transforms the
$V(E,r)$ for each chemical potential $\mu$ and temperature $T$, to an
equivalent local potential (ELP) $V_L(r)$ which is energy-independent.
The construction of the $V_L(r)$ is thus desirable for two reasons.
Firstly, to investigate how the characteristics of the in-medium
two-body interaction are manifested in the presence of the Fermi
function $f(E)$.  Secondly, to enable us to employ a configuration
space formalism describing the A-body system embedded in nuclear
matter.\par

In this respect, the two-variable integro-differential equation 
formalism (IDEA) in configuration space \cite{IDEA1,IDEA2,O16}, 
valid for three- and  many--body systems is employed. The IDEA 
is based on the Faddeev formalism, where the A-body 
wave function can be written as a sum of two-body amplitudes, and
on the Hyperspherical Harmonics.  The reduced equations
give binding energy results for the A-body   which are in excellent 
agreement with  variational, Green Function Monte Carlo,
and other competing methods.\par

In what follows, we shall describe first in Sect. \ref{sec:form}
our formalism  starting from a brief description of the effective 
in-medium equations,  the localization procedure, and the 
IDEA method. Our results are discussed in Sect. \ref{sec:res}
and our conclusions are summarized in Sect. \ref{sec:con}

\section{Formalism}\label{sec:form}

We use the techniques of many-body Green functions.  Reference to
textbook treatments is given in~\cite{fet71}. The Dyson equation
approach to correlations along with several approximation schemes has
been developed in~\cite{sk71,sk73,sk90,sk91,sk94,sk96}, reviewed
in~\cite{duk98} and will be briefly summarized in the next section.
In recent years we have systematically applied this approach up to
four-nucleon
clusters~\cite{Beyer:1996rx,bey97,Beyer:1997sf,Beyer:1999tm,Beyer:1999zx,Kuhrts:2000jz,Beyer:1999xv,Beyer:2000ds,Kuhrts:2001zs,Beyer:2003zz}
and also generalized it to the light front to describe quark
correlations in hot and dense relativistic nuclear
matter~\cite{Beyer:2001bc,Mattiello:2001vq,Beyer:2003qb,Beyer:2003ca}.

\subsection{The Dyson equation approach}
\label{sec:Dyson}
Let the many-particle Hamiltonian be given by
\begin{eqnarray}
    H =\sum_{11'} H_0(1,1')a_1^\dagger a_{1'}+\;\frac{1}{2}\;
    \sum_{12 1'2'}V_2(12,1'2')\;a^\dagger_1  a^\dagger_2  a_{2'} a_{1'}
\end{eqnarray}
where $H_0$ is the kinetic energy, $V_2$ a generic two-body potential,
and $a$, $a^\dagger$ are Fermionic destruction and creation operators.
 We define a chronological
Green function for a given number of particles as follows
\begin{eqnarray} 
      {\cal G}_{\alpha\beta}^{t-t'}&=&-i\langle T A_\ga(t)
          A_\gb^\dagger(t')\rangle\nonumber\\
       &=&-i\left(\theta(t-t')\langle A_\gb^\dagger(t')A_\ga(t)
      \rangle\mp\theta(t'-t)\langle A_\gb^\dagger(t')A_\ga(t)
       \rangle\right).
\label{eqn:green}
\end{eqnarray}
The average $\langle\cdots\rangle$ is taken over the exact ground
state and the upper (lower) sign is for fermions (bosons). The operators
$A_\ga(t)$ could be build out of any number of destruction and/or
creation operators. For $n$-particle correlations
$A_\ga=a_1a_2a_3\dots a_n$.  The time dependence of the operators is
given in the Heisenberg picture by $A(t)=e^{iHt}Ae^{-iHt}$. Note that
the multi-particle (or cluster) Green function is defined at equal or
global time.

In finite temperature formalism the above definition can be
generalized.  For a grand canonical ensemble, relevant here,
the Heisenberg picture assumes the form $A(\tau)=e^{(H-\mu
  N)\tau}Ae^{-(H-\mu N)\tau}$, which includes the chemical potential
$\mu$ and the number operator $N$. For $\tau=it$ (imaginary time) the
original Heisenberg picture is formally recovered. The average is now
taken over the  (equilibrium) grand canonical statistical operator
$\rho$, viz.  $\langle\cdots\rangle={\rm tr}\{\rho\dots\}$. To
accommodate the convention of Fetter and Walecka~\cite{fet71} the
imaginary time Green function is given by
\begin{equation} 
    {\cal G}_{\alpha\beta}^{\tau-\tau'}
        =-\langle T_\tau A_\ga(\tau)A_\gb^\dagger(\tau')\rangle,
\end{equation}
where the time ordering is according to the value of $\tau$, with the
smallest at the right and the imaginary unit is dropped compared to
(\ref{eqn:green}).

Dyson equations can be established for both forms~\cite{duk98,bey97},  
\begin{equation} 
    i\frac{\partial}{\partial\tau}\; {\cal G}^{\tau-\tau'}_{\ga\gb}
     =\delta(\tau-\tau') \langle [A_\ga,A_\gb^\dagger]_\pm\rangle
    + \sum_{\gc}\int d\bar \tau\; {{\cal M}^{\tau-\bar \tau}_{\ga\gc}}
    \;{\cal G}^{\bar \tau-\tau'}_{\gc\gb}.
\label{eqn:dyson}
\end{equation}
The mass matrix that appears in (\ref{eqn:dyson}) is given by
\begin{eqnarray} 
        {\cal M}^{\tau-\tau'}_{\ga\gb}
        &=& \delta(\tau-\tau') {\cal M}^{\tau}_{0,\ga\gb}
        + {\cal M}^{\tau-\tau'}_{r,\ga\gb}\label{eqn:mass}\\
       ( {\cal M}_0{\cal N})^\tau_{\ga\gb}&=
         &\langle [[A_\ga,H](\tau),A_\gb^\dagger(\tau)]\rangle
\label{eqn:mass0}\\
        ({\cal M}_r{\cal N})^{\tau-\tau'}_{\alpha\beta}&=&
       \sum_\gamma\langle T_\tau [A_\alpha,H](\tau),
       [A^\dagger_\gb,H](\tau')]\rangle_{{\rm irreducible}}
\end{eqnarray} 
where we have used 
${\cal N}^\tau_{\ga\gb}=\langle[A_\ga,A_\gb^\dagger]_\pm(\tau)\rangle$.
The first term in (\ref{eqn:mass}) is instantaneous and related to the
mean field approximation, the second term is the retardation term.
This form suggests to first solve the mean field problem (neglecting
the retardation term) and evaluate higher order contributions
involving ${\cal M}^{\tau-\tau'}_{r,\ga\gb}$ in the mean field basis.
Hence (\ref{eqn:dyson}) may be written in the following form
\begin{eqnarray} 
       i\frac{\partial}{\partial\tau}\; {\cal G}^{\tau-\tau'}_{0,\ga\gb}
        &=&\delta(\tau-\tau') {\cal N}^\tau_{\ga\gb}
      + \sum_{\gc}\; {{\cal M}^{\tau}_{0,\ga\gc}}
      \;{\cal G}^{\tau-\tau'}_{0,\gc\gb},\label{eqn:mf}\\
       {\cal G}^{\tau-\tau'}_{\ga\gb}
       &=&{\cal G}^{\tau-\tau'}_{0,\ga\gb}
      + \sum_{\gc\gd}\int\int d\bar \tau d\bar \tau'\; 
      {\cal G}^{\tau-\bar\tau}_{0,\ga\gc}
      ({\cal N}^{-1}{\cal M}_r)^{\bar\tau-\bar \tau'}_{\gc\gd}
       \;{\cal G}^{\bar \tau'-\tau'}_{\gd\gb}.\label{eqn:red}
\end{eqnarray}
%
As a first approximation, we neglect the retardation part and hence solve
only the mean field equation. 

It is convenient to use the Matsubara-Fourier
representation~\cite{fet71}
of the Dyson equations. This leads to the Matsubara frequency instead
of the ``time'' derivative in (\ref{eqn:mf})
\begin{equation}
     z_\gl = \frac{\gl\pi}{-i\gb}+\mu
\end{equation}
where $\gl=2n$ for bosons and $\gl=2n+1$ for fermions. 

For a one body operator, (\ref{eqn:mf}) leads to the standard mean
field approximation (Hartree-Fock approximation) and an effective
Hamiltonian that describes uncorrelated quasiparticles. To
be more specific, for $A_\ga=a_1$ the one-particle Green function,
 as follows from (\ref{eqn:mf}), the use of (\ref{eqn:mass0})
 and the of Matsubara-Fourier transformation,  is
\begin{equation}
          G(z)=(z-\varepsilon_1)^{-1}
\end{equation}
where the Matsubara frequency has been analytically continued
$z_\gl\rightarrow z$. The quasiparticle self energy is
\begin{equation}
       \varepsilon_1 = \frac{k^2_1}{2m_1}+\sum_{2}V_2(12,
       \widetilde{12})f_2 = \frac{k^2_1}{2m_1}+\Delta^{\rm HF}(1)
\label{eqn:eps}
\end{equation}
where $\widetilde{12}$ denotes proper antisymmetrization of particles
1 and 2.  The Fermi function $f_i\equiv
f(\varepsilon_i)$ for the $i$-th particle in an uncorrelated medium is
given by
\begin{equation}
     \langle a^\dagger a\rangle = {\rm tr}(\rho a^\dagger a)
      =f(\varepsilon_i) = \frac{1}{e^{\gb(\varepsilon_i - \mu)}+1}\, .
\end{equation}
The density operator of an uncorrelated medium is characterized only
by single particle operators, viz.
\begin{equation}
     \rho_0=\frac{e^{-\gb K}}{{\rm tr}(e^{-\gb K})},\qquad  
      K=\sum_i (\varepsilon_i-\mu) a^\dagger_i a_i.
\end{equation}
For a general operator $A_\ga$ other than one body, 
(\ref{eqn:mf}) is called cluster mean field approximation. Using the
density operator for an uncorrelated medium to evaluate the traces,
the equation hierarchy decouples.
This is a consistent way to arrive at equations for each cluster that
can be solved with few-body techniques.

The resolvent $G_0$ for $n$ noninteracting quasiparticles is
\begin{equation}
      G_0(z) = (z -  H_0)^{-1}
      N \equiv R_0(z) N,\qquad H_0 = \sum_{i=1}^n \varepsilon_i
\end{equation}
where $G_0$, $H_0$, and $N$ are formally matrices in $n$ particle
space. The Pauli-blocking for $n$-particles is
\begin{equation}
       N=\bar f_1\bar f_2 \dots \bar f_n
           \pm f_1f_2\dots f_n,\qquad\bar f=1-f
\end{equation}
Note that $NR_0=R_0N$. Straightforward but tedious evaluation of
(\ref{eqn:mf}) using $A_\ga=a_1a_2a_3\dots a_n$ (\ref{eqn:dyson}) and
the Wick theorem, leads also to the full resolvent $G(z)$. For each
number of particles in the cluster the resolvents
have the same formal structure and after Matsubara-Fourier
transformation can be written in a convenient way
as in the isolated system, viz.
\begin{equation}
     G(z)=(z-H_0-V)^{-1}{N}\equiv R(z){N}, \qquad 
       V\equiv \sum_{\mathrm{pairs}\;\ga} N_2^{\ga}V_2^{\ga}\,.
\end{equation}
Note that $V^\dagger\neq V$ and $R(z)N\neq NR(z)$. To be specific,
for an interaction in pair $\ga=(12)$ the effective potential
reads 
\begin{equation}
     \langle 123\dots|{N_2^{(12)}}V_2^{(12)}|1'2'3'\dots\rangle = 
     {(\bar f_1\bar f_2 - f_1f_2)}V_2(12,1'2')\gd_{33'}\dots
\end{equation}
For further use in the Alt-Grassberger-Sandhas (AGS) 
equations~\cite{ags1,ags2}, we give also the channel resolvent
\begin{equation}
      G_\ga(z)=(z-H_0-  N_2^{\ga}V_2^{\ga})^{-1} N\equiv R_\ga(z) N.
\end{equation}
For the two-body case as well as for a two-body subsystem embedded in
the $n$-body cluster the standard definition of the $t$ matrix leads
to the familiar $t$ matrix  equation for finite temperature and
densities~\cite{fet71},
\begin{equation}
    T_2^\ga(z) =   V_2^\ga +  V_2^\ga  N^\ga_2 R_0(z)  T_2^\ga(z).
\label{eqn:T2}
\end{equation}
This equation,  given in the quasiparticle approximation, is used as
a driving kernel for the few-particle equations. Going beyond the 
quasiparticle approximation would extend the scope of the present 
work. Note, however, that improvements can be systematically
achieved by including the retardation part of the effective mass
matrix, although technically this is  quite a difficult task. If
multi-particle correlations are neglected (binary collision 
approximation) it is possible to solve the $t$
matrix in a self consistent way, see e.g. \cite{Bozek:1999su}.

\subsection{Effective in-medium Equations}
\label{sec:IME}
The in-medium Schr\"odinger type equation equivalent to (\ref{eqn:T2})
is given by
\begin{equation}
       (H_0  - z) \Psi(12)  +
             \sum_{1'2'} (1-f_1-f_2) V_2(12,1'2') \Psi(1'2') = 0.
\label{eqn:S}
\end{equation}
In the present work we are interested in a region of rather low
density in the vicinity of the Mott transition ($\sim \rho_0/10$,
where $\rho_0=0.16$ fm$^{-3}$ is normal nuclear matter density). 
We therefore use an effective mass approximation for the quasi-particle
energy $\varepsilon$; {\em i.e.}
 after evaluation of Eq.~(\ref{eqn:eps}) we fit the
effective mass $m_{\mathrm{eff}}$ via
\begin{equation}
     \varepsilon=\frac{k^2}{2m}+\Delta^{\rm HF}(k)
        \simeq\frac{k^2}{2m_{\mathrm{eff}}}+\Delta^{\rm HF}_0.
\end{equation}
The constant shift $\Delta^{\rm HF}_0$ can be absorbed in a redefinition
of the chemical potential $\mu_{\rm eff}=\mu-\Delta^{\rm HF}_0$.  
To simplify our analysis further, we assume that the two-body system 
is at rest in the medium. It is well known that the influence 
of the medium fades away for larger relative momentum respective 
to the medium \cite{schmidt:1990}.  Upon introducing relative 
and c.m.  coordinates for the two-particle system ($p$ and $P=0$), 
the kinetic energy term reads
\begin{equation}
      \frac{k_1^2}{2m_{\mathrm{eff}}}=\frac{k_2^2}{2m_{\mathrm{eff}}}
             =\frac{p^2}{2m_{\mathrm{eff}}}=\frac{1}{2}\, E
\end{equation}
and $E=z-E_{\mathrm{cont}}$ is the two-body binding/scattering 
energy in the c.m. system. In the effective mass approximation and
for  $P=0$,  $E_{\mathrm{cont}}$ is given by 
$E_{\mathrm{cont}}=2\Delta^{\rm  HF}_0$\cite{schmidt:1990}.
  Using a local bare nucleon nucleon
potential $V_2(r)$, where $r$ is the conjugate of $p$, the
effective potential becomes energy dependent
\begin{equation}
          V(E,r)=(1-2f(E))\;V_2(r), 
\label{vef}
\end{equation}
where the Fermi function is now explicitly given by
\begin{equation}
         f(E)=\frac{1}{e^{\beta(E/2 - \mu_{\mathrm{eff}})}+1}.
\end{equation}
The effective in-medium Schr\"odinger-type equation for 
scattering states then reads
\begin{equation}
          \left(\frac{p^2}{m_{\mathrm{eff}}} 
                   + V(E,r) - E\right)\;\psi(E,r) = 0.
\end{equation}
Note that this equation holds for $E>0$, since
$E=p^2/m_{\mathrm{eff}}$ for the eigenvalues.\par 

The  potential $ V(E,r)$  can be employed to study its 
effects and how is manifested in $A$-body systems. For
example, for $A=2$, one may study its analytical behavior
in the complex $k$-plane as the  nuclear density $\rho$ changes
\cite{BS01} or study the influence it has on clusters
embedded in the medium. Due to its  energy-dependence,
it can be employed in momentum space integral equations for three-- 
\cite{Beyer:1999zx} and four--body  \cite{Beyer:2000ds} systems 
embedded in the medium.  \par

\subsection{Energy-independent interactions}
Equivalence between two potentials  $V_1$ and $V_2$, irrespective
of their nature and structure, means that the scattering
wave functions for the two-potentials are related by \cite{FS83}
\begin{equation}
         u_1(r)=f(r)u_2(r)
\label{elp}
\end{equation}
with
\begin{equation}
 \qquad          f(r)\mathop{\longrightarrow
      }_{ r \rightarrow \infty}  1
\label{fasym}
\end{equation}
{\em i.e} the two wave functions must be asymptotically 
the same and thus the phase shifts are also the same. In our case, 
however, we have to obtain an equivalent potential to $V(E,r)$  
which should  not only provide the same asymptotic 
scattering wave function but it should generate differences  
in the interior region which are minimal 
($f(r) \sim 1 \ \forall\ r$) while at the same time 
the corresponding binding energy $E_b$ obtained by the
$E$-independent interaction should be the same 
as the one obtained from $V(E_b,r)$.\par

From all methods  employed to construct $E$--independent
ELP's,  the inverse scattering technique is the most natural one. 
The two-body potential is, 
in this case, directly obtained from the available bound and scattering 
information in a unique way and without
an  {\it a priori} assumption about the shape and range of the interaction.
This not only guarantees the recovery of the on-shell behavior
of the wave function but of the bound states -when present- as well.
The off-shell differences between the original  wave function
and the one generated by the ELP potential obtained by inversion,
described by the function $f(r)$, Eq. (\ref{fasym}), are also
minimal.    From all variant inverse scattering methods, the  fixed-$\ell$
inversion of Marchenko \cite{Mar,Chadan,Newton} is perhaps the most 
suitable to employ in order to construct an $E$-independent 
ELP to $V(E,r)$ as defined by (\ref{vef}). The required  
input is the $S$-matrix $S_\ell(k)$, defined by
\begin{equation}
         S_\ell(k)=\frac{f_\ell(-k)}{f_\ell(k)}={\rm e}^{2i\delta_\ell(k)}\,,
\label{smatr}
\end{equation}
where $f_\ell(k)$ is the Jost-function the analytical properties
of which describe the scattering, bound, and resonances
of the system and $\delta_\ell(k)$ are the scattering phase shifts
at the energy $E=\hbar^2k^2/2\mu$    which   
are used to obtain  the Fourier transform  ${\cal F}_\ell(r,r')$ of the 
$S$-matrix $S_\ell(k)$,
\begin{eqnarray}
     {\cal F}_\ell(r,r')&=&\frac{1}{2\pi} \int_{-\infty}^\infty
     h_\ell^{(+)}(kr)\left[1-S_\ell(k)\right]h^{(+)}_\ell(kr')
      {\rm d}k\nonumber\\&&+
      \sum_{n=1}^{N_b}
     A_{n \ell}h^{(+)}_\ell(b_n r)h^{(+)}_\ell(b_n r')\ ,
\label{frr}
\end{eqnarray}
where $h^{(+)}_\ell(z)$ is the Riccati-Hankel function, $N_b$ is the
number of bound states, and $A_{n \ell}$ are the corresponding
asymptotic normalization constants \cite{Newton}.  
The ${\cal F}_\ell$ in turns serve as an input 
to the  Marchenko integral equation, 
\begin{equation}
        K_\ell(r,r')+{\cal F}_\ell(r,r')+\int_r^\infty
                K_\ell(r,s){\cal F}_\ell(s,r'){\rm d} s=0\ .
\label{March}
\end{equation}
Once the $K_\ell(r,r')$ is obtained from the solution of 
(\ref{March}), the potential $V_\ell(r)$ for the partial wave 
$\ell$ is calculated from the relation
\begin{equation}
        V_\ell(r)=-2\;\frac{{\rm d} K_\ell(r,r)}{{\rm d}r}.
\label{Vr}
\end{equation}
It is clear that while in the Schr\"odinger equation 
one starts from the knowledge of the potential to
obtain the  wave function which provide us with
 the physical information, in the inverse scattering procedure 
one uses instead experimental scattering data
(or data obtained via a theoretical approach)  to construct 
the underlying  unknown potential first.
The practical implementation of the method has been described
in \cite{BS01,Alt94}.\par

\subsection{Integro-differential Equations Approach}
The method we employ to study the dynamics of the
A-boson\label{idea}\footnote{Presently we neglect spin-isospin
  dependence, which means that the effective few-body equations
  derived are for bose-like particles. This restriction seems not
  severe as it might be interpreted as a spin averaging. This kind of
  averaging procedures (e.g. angular momentum averaging of Pauli
  factors) are widely used for quasiparticles and lead for example to
  the possibility of angular momentum expansion. This does not affect
  the Fermi statistics in use.}  like system is based on the Faddeev
decomposition of the wave function and on Hyperspherical Harmonics.
The $A$-body Schr\"odinger equation reads
\begin{equation}
       \bigg [T+\sum_{i<j\le A} V(r_{ij})-E)\bigg]\Psi({\bf x})=0
\label{SchrA}
\end{equation}
 where ${\bf x}$ stands for all the coordinates of all particles in the 
center-of mass frame, $ V(r_{ij})$ is a central two-body potential, and
 $r_{ij}=|{\bf x}_i-{\bf x}_j|$. Assuming that the wave function
is written as a sum of amplitudes, 
\begin{equation}
       \Psi({\bf x})=\sum_{i<j\le A}\Phi({\bf r}_{ij},{\bf x})
      =H_{[L_m]}({\bf x}) \sum_{i<j\le A}    F({\bf r}_{ij},r)
\label{psiex}
\end{equation}
where $H_{[L_m]}({\bf x})$ is a harmonic polynomial of minimal degree $L_m$
and $r$ is the hyperradius $r=[(2/A)\sum_{i<j\le A} r^2_{ij}$, 
one obtains the Faddeev-type equation for the amplitude 
$F({\bf r}_{ij},r)$
\begin{equation}
       \big [T-E\big]\,H_{[L_m]}({\bf x})F({\bf r}_{ij},r)=-V(r_{ij})
        H_{[L_m]}({\bf x})\sum_{k<l\le A}F({\bf r}_{kl},r)
\label{FanA}
\end{equation}
This equation may be modified by introducing in both sides 
the average of the potential over the unit hypersphere {\em i.e}
the so-called hypercentral potential $V_0(r)$ defined by \cite{IDEA1}
\begin{equation}
        V_0(r)=\int_{-1}^{+1}V(r\sqrt{(1+z)/2})W(z) {\rm d}z
\label {V0}
\end{equation}
where $z=2 r^2_{ij}/r^2-1$ and where the weight function $W(z)$  is given
by
\begin{equation}
      W(z)=(1-z)^\alpha(1+z)^\beta
\label {Wz}
\end{equation}
with $\alpha=(D-5)/2$, $\beta=1/2$, and  $D=3(A-1)$.
Then  instead of (\ref{FanA}) one has the modified Faddeev
equation
\begin{eqnarray}
\nonumber
      &&\hspace*{-1.5cm} \bigg [T+\frac{A(A-1)}{2}V_0(r)-E)\bigg]H_{[L_m]}({\bf x})
      \phi({\bf r}_{ij},r)
\\ 
   &&\hspace*{1.5cm}=-\bigg[V(r_{ij})-V_0(r)\bigg]
        H_{[L_m]}({\bf x})\sum_{k<l\le A}\phi({\bf r}_{kl},r)
\label{FanB}
\end{eqnarray}
It is noted that by summing aver all pairs $(ij)$ one recovers
the Schr\"odinger equation. Letting 
$\phi({\bf r}_{ij},r)=P(z,r)/r^{(D-1)/2}$ for the modified 
Faddeev component, 
 multiplying from left by $H^*_{[L_m]}({\bf x})$,
 and integrating over the $(N-1)$-hyperangles ${\rm d}\Omega_{N-1}$ 
(except those of the $(i,j)$-pair) we obtain 
the integro-differential equation (IDEA) for the amplitude
$P(z,r)$ \cite{IDEA1,IDEA2} 
\begin{eqnarray}
\nonumber
    \bigg\{\frac{\hbar^2}{m}\bigg [
    {\rm D}_r+\frac{4}{r^2}{\rm D}_z \bigg]
        &+&\frac{A(A-1)}{2}V_0(r) -E \bigg\}\, P(z,r)
\\
     &=&-\left[ V(r\sqrt{(1+z)/2})-V_0(r)\right] \Pi(z,r)  
\label{idea}
\end{eqnarray}
where we use the abbreviations
\begin{equation}
\label{Drz}
     {\rm D}_r\equiv - \frac{\partial^2}{\partial r^2}+
              \frac{{\cal L}({\cal L}+1) }{r^2}\,,
\qquad
         {\rm D}_z =-\frac{1}{W(z)}\frac{\partial}{\partial z}(1-z^2)
          W(z)\frac{\partial}{\partial z}
\end{equation}
and
\begin{equation}
         \Pi(z,r)=P(z,r)+\int_{-1}^{+1}f(z,z') P(z',r)\,{\rm d}z'\,.
\end{equation}
In the above,  ${\cal L}=(D-3)/2$ and the function
$f(z,z')$ is given  by
\begin{equation}
        f(z,z')=W(z')\sum_K(f_k^2-1)P_K(z)P_K(z')/h_K
\label {fzz}
\end{equation}
where $P_K(z)\equiv P^{\alpha,\beta}_K(z)$ are Jacobi
polynomials associated to the weight function $W(z)$.
  The constant $f_K$ is given by 
\begin{equation}
        f_K^2=1+\left [ 2(A-2)P_K(-1/2)+\frac{(A-2)(A-3)}{2}P_K(-1)
              \right]/P_K(1)
\label {fK}
\end{equation}
while  $h_K$, a  normalization constant, by
\begin{equation}
        h_K=\int_{-1}^{+1} \left[ P_K(z)\right]^2W(z){\rm d}z
\label {hK}
\end{equation}
Details on the IDEA formalism can be found in Refs. \cite{IDEA1,IDEA2} 
and will not be discussed here. \par

In the adiabatic approximation one assumes
that the radial motion (coordinate $r$) and the orbital motion 
(coordinate $z$) are nearly decoupled
{\em i.e} we may right
    $ P(z,r)\approx P_\lambda(z,r)u_\lambda(r)$
and that   $P_\lambda(z,r)$ is varied slowly with $r$ and thus  one can omit
its derivatives with respect to  $r$. Then Eq. (\ref{idea}) 
can be split into two equations ($\hbar^2/m=1$)
\begin{equation}
     \left [-\frac{4}{r^2}{\rm D}_z
        +U_\lambda(r)\right]\,P_\lambda(z,r)
       =\left[ V(r\sqrt{(1+z)/2})-V_0(r)\right] \Pi_\lambda(z,r)  
\label{eaa1}
\end{equation}
and 
\begin{equation}
\label{eaa2}
      \left[ D_r+\frac{A(A-1)}{2}V_0(r)+U_\lambda(r))-E_\lambda^{{\rm eaa}}
         \right]u_\lambda(r)=0
 \end{equation}
The $U_\lambda(r)$ is the so-called eigen-potential sometimes referred to
as potential surface.  It is a characteristic for each
system  and it can be used to calculate  reliably the 
$A$-boson like bound states via (\ref{eaa2}). The solution thus 
obtained provide us a lower bound for the binding energy
  $E_\lambda^{{\rm eaa}}$ \cite{IDEA1,IDEA2,O16}.
\par

\section{Results}\label{sec:res}
In our investigations we employ the Volkov potential \cite{Volkov} 
which is widely used in model nuclear structure calculations,
\begin{equation}
\label{volkov}
      V(r)=-83.3400196\exp(-(r/1.6)^2)+144.843409\exp(-(r/0.82)^2)\,
\end{equation}
the two-body binding energy being $E_2=0.5462$\,MeV.
The medium-modified energy dependent potentials $V(E,r)$
are constructed  using Eq. (\ref{vef}). Employing  $V(E,r)$
one can  obtain phase shifts at any energy $E$ corresponding to 
a specific  $\mu_{\mathrm{eff}}$ and  effective mass $m_{\mathrm{eff}}$. 
In the present work we use two nuclear densities
namely  $\rho=0.003$ fm$^{-3}$ and $\rho=0.034$ fm$^{-3}$  and $T=10$. 

\begin{figure}[b]
\vspace*{-8mm}
\begin{minipage}[t]{7cm}
\includegraphics[width=6.8cm,angle=360]{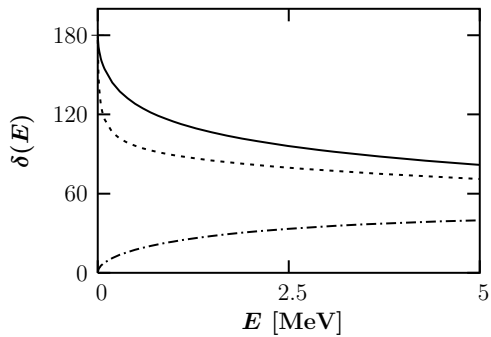}
\end{minipage}
\vfill\hfill
\begin{minipage}[t]{5cm}
\caption{\label{del} 
The low energy behavior of the phase shifts
without distortion (---), for  $\rho=0.003$\, fm$^{-3}$ (-\,-\,-),
and for $\rho=0.034$\, fm$^{-3}$ (-\,$\cdot$\,-)
at $T=10$\, MeV. 
The latter phase shifts start from zero implying the absence 
of a bound state.}
\end{minipage}
\vfill
\vskip 1.6cm
\end{figure}

The low energy behavior of these phase shifts are shown in Fig. \ref{del}.
The distortion of the potential for $\rho=0.003$ fm$^{-3}$ 
results in different phase shifts but the  potential still  
sustains a bound state at $E_2=0.0282$\, MeV.
In contrast the phase shift for $\rho=0.034$ fm$^{-3}$
start from zero and thus, according to Levinson's theorem, no bound 
exists in this case.\par

\begin{figure}[t]
\hspace*{-.2cm}
\includegraphics[width=6.25cm,angle=360]{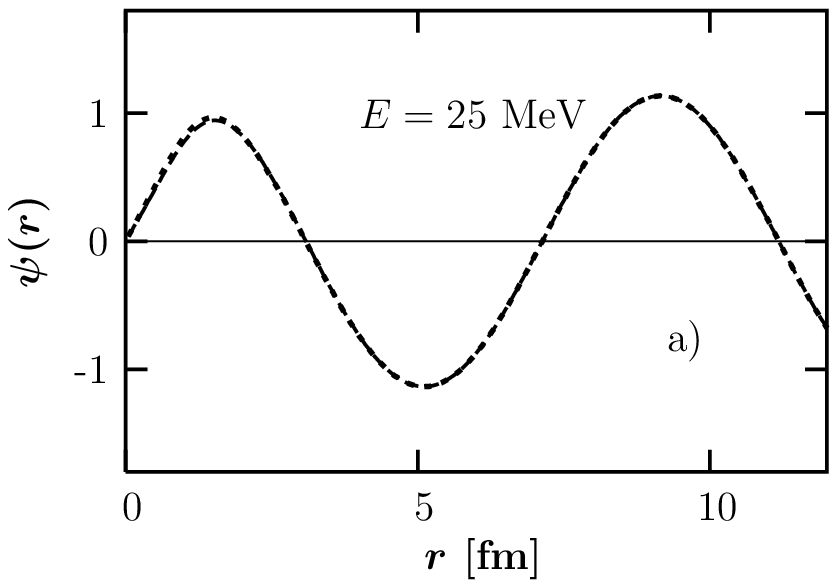}
\vfill
\vspace*{-4.67mm}
\hspace*{6cm}
\includegraphics[width=6.25cm,angle=360]{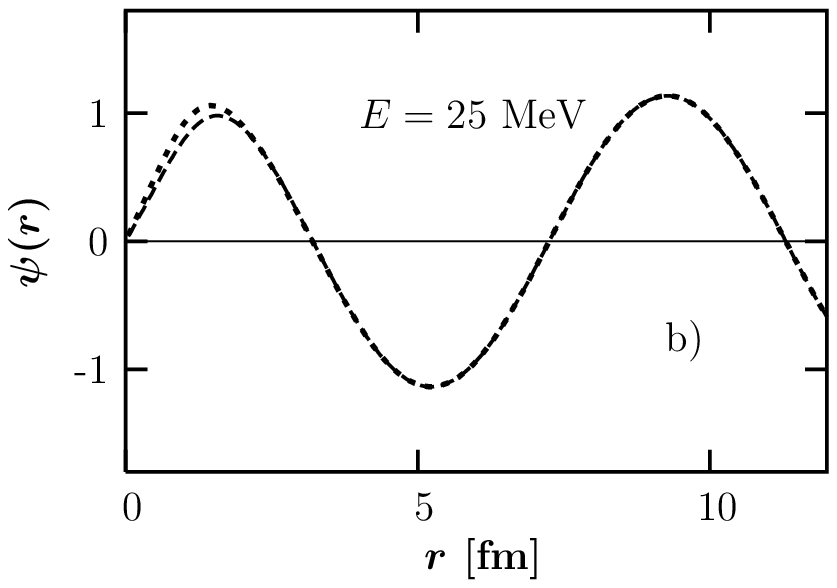}
\vspace*{4.5cm}
\caption{\label{wfs}
  Comparison of the scattering wave functions $\psi(r)$ at 
$E=25$\,MeV obtained with the $E$-independent Marchenko 
  potential  $V_{\rm L}(r)$ (---) with
  those of $V(E,r)$ (-\,-\,-) for $\rho=0.003$\, fm$^{-3}$ (Fig. a) )
and  $\rho=0.034$\, fm$^{-3}$ (Fig. b) )
at $T=10$\, MeV.} 
\vspace*{-4mm}
\end{figure}

Using the above phase shifts we constructed the $E$-independent
potentials by means of the Marchenko inverse scattering method,
the quality of which can be seen in Fig. \ref{wfs} where
the original and the resulted wave functions at $E=25$\,MeV 
are drawn. It is seen that, to all practical purposes, the
two wave functions are the same {\em i.e} the
 potential $V(E,r)$  is not only on--shell equivalent 
but (nearly) off--shell equivalent to the one obtained by inversion,
 $V_{\rm L}(r)$, as well. Similar results are obtained at all energies.\par
%

\begin{figure}[ h]
\hspace*{.1cm}
\includegraphics[width=6.25cm,angle=360]{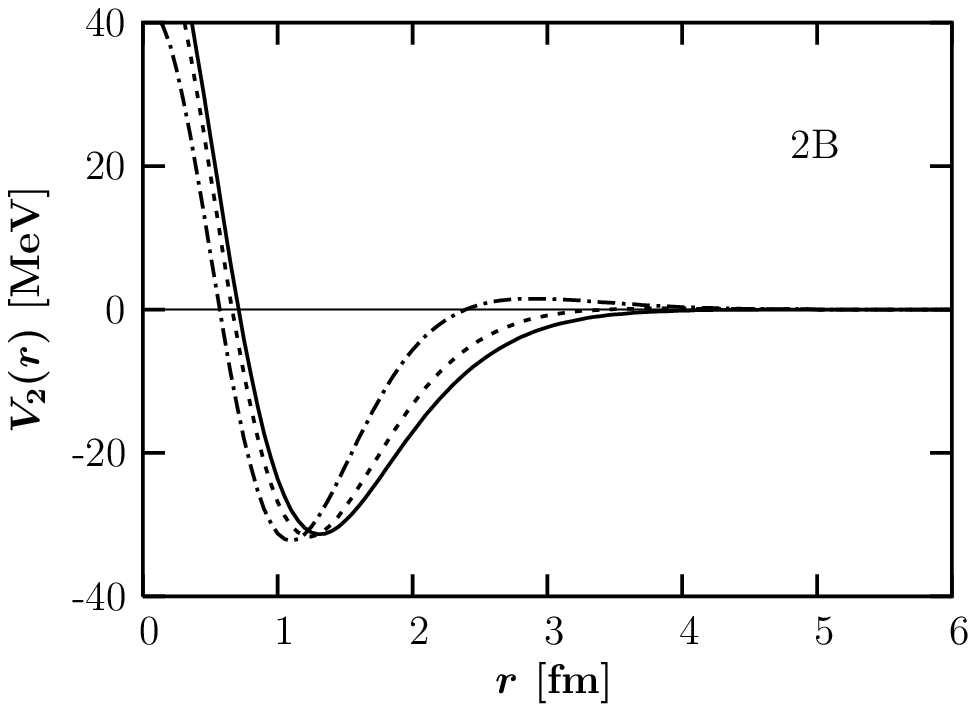}
\vfill
\vspace*{-4.67mm}
\hspace*{6.2cm}
\includegraphics[width=6.25cm,angle=360]{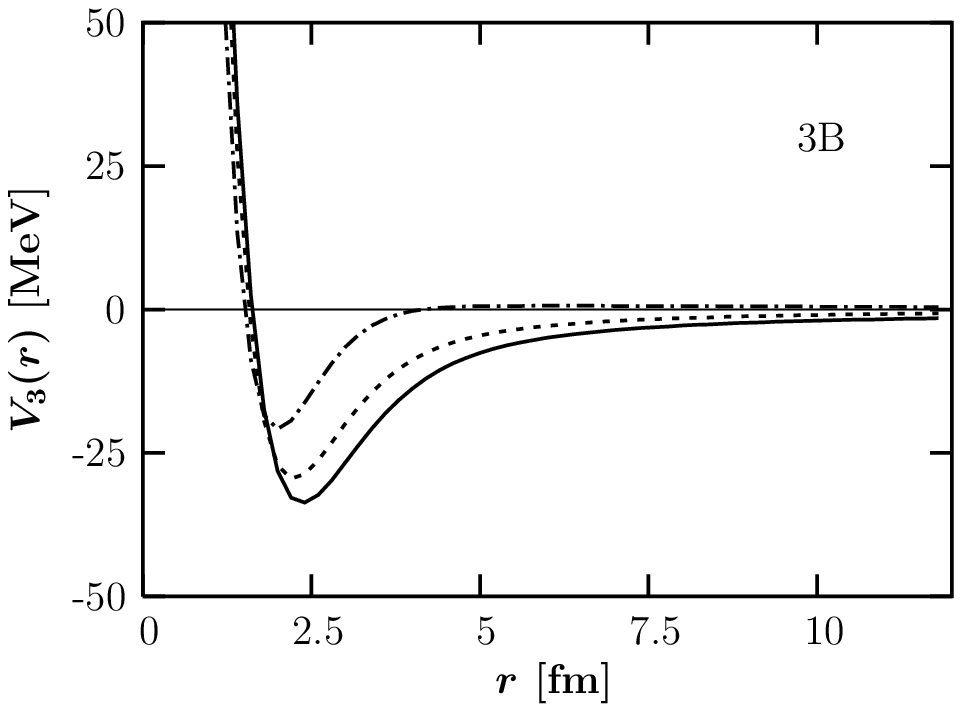}
\vspace*{4.cm}

\hspace*{.0cm}
\includegraphics[width=6.2cm,angle=360]{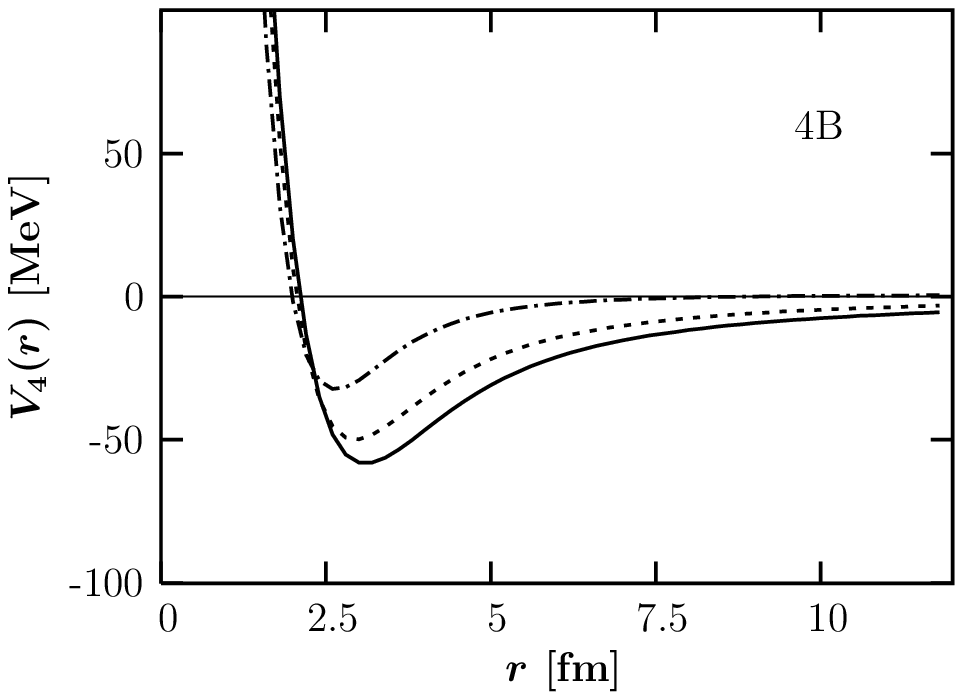}
\vfill
\vspace*{-11.8mm}
\hspace*{6cm}
\includegraphics[width=6.5cm,angle=-360]{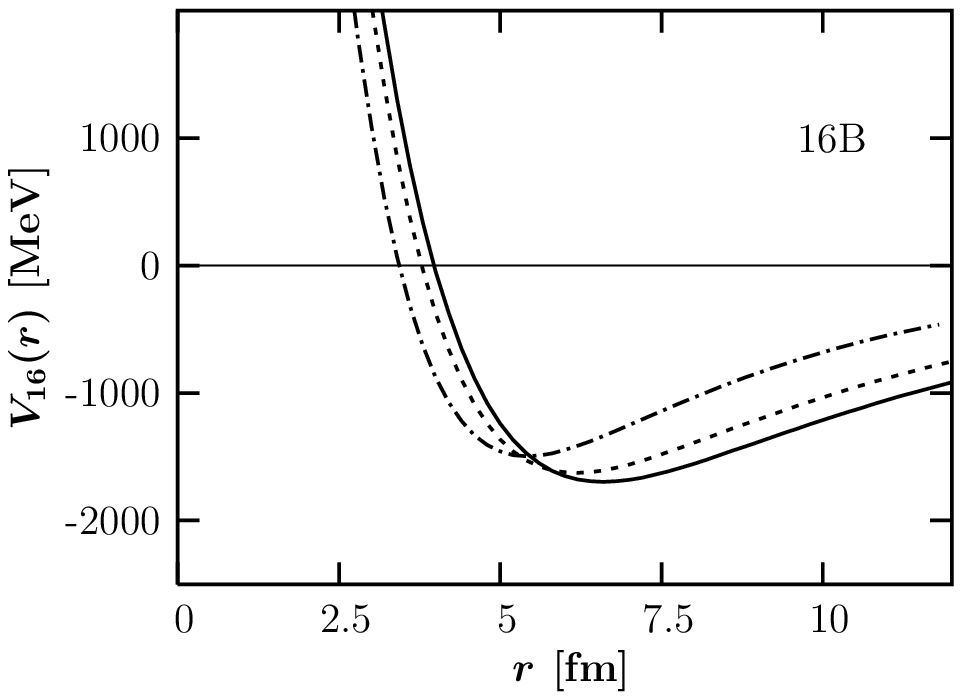}

\vspace*{4.6cm}
\label{vs}
\caption{
  The two-body potentials and the corresponding  potential energy 
surfaces (in the extreme adiabatic approximation) for 3-, 4- and 16-body
systems for zero distortion (- - -),  for $\rho=0.003$\, fm$^{-3}$
  (-\,-\,-), and  $\rho=0.034$\, fm$^{-3}$ (-\,$\cdot$\,-).}
\end{figure}

Once the quality of the interactions is checked, they are 
employed  in the IDEA equations (\ref{eaa1}) and (\ref{eaa2}) 
to obtain the binding energies for the three--, four--
and sixteen--body systems in the adiabatic approximation.  
The binding energy results for the three
densities used are given in table \ref{Eb}. It is seen
that  with the bare interaction used,   the three-body
 is the first to 'dissolve' in the medium for  density 
$\rho=0.034$ fm$^{-3}$. The four-body results show 
similar trends to the three-body system. However,  the binding energy 
for the  16--body system,  is not as drastically reduced and
for $\rho=0.034$ fm$^{-3}$ which is quite higher
than the Mott density, the binding energy is still significant.
\begin{table}[hbt]
\begin{center}
\caption{\label{Eb}
Binding energy results for 2--, 3--, 4--, and 16--body systems.}
\begin{tabular}{lllll}
\hline
$\rho$ (fm$^{-3}$) & $E_2$ (MeV) & $E_3$ (MeV) &$E_4$ (MeV) &$E_{16}$ (MeV)\cr
\hline
    0     &  0.5462     & 8.6704   & 30.7250    & 1630.1462   \cr
    0.003 & 0.0282      & 4.7202   & 22.2798    & 1554.0695   \cr
    0.034 &   -         &    -     & 5.4093     &  1410.5321  \cr
\hline
\end{tabular}
\end{center}
\end{table}
In  Fig. \ref{vs} the ELP's for the medium modified 2-body
interaction is plotted together with the potential surfaces  
obtained from the solution  of (\ref{eaa1}). It is seen that 
the essential features of the medium modified 
two-body interaction with the Volkov (bare) potential are
similar to those obtained Ref. \cite{BS01} with the Malfliet-Tjon 
potential: As the density of the medium increases the 
attractive well becomes narrower
and eventually for densities above the Mott one, the two nucleons 
become unbound. It is interesting to note that at these densities
the repulsion in the interaction region appears. The analytical
properties for the corresponding Jost function  of the potentials 
have been  studied in Ref. \cite{BS01} where it was 
shown that the bound state, instead of being
continued off the imaginary axis to the resonances region, 
moves along the negative imaginary $k$-axis, i.e.  it becomes an 
{\em anti-bound} state -- at least for the densities used where
the repulsive hump in the interaction region is weak.
The potential surfaces exhibit similar behavior and in addition 
they become shallower.
The eigen-potentials  for the 16--body system 
tend, in addition, to shift the repulsion  in the 
interior region  in a more pronounced manner.

\section{Conclusions}\label{sec:con}
We propose a method to investigate the A-body system embedded in the 
medium   at a finite temperature and for densities   below and 
above the Mott density. Unlike previous investigations 
which were carried out in momentum space, the present one
is in configuration space and thus advantageous in  visualizing
the effects of the medium on the system under consideration.  \par

To achieve this, we firstly apply inverse scattering
techniques to transform the medium modified
 two-body interactions which are energy dependent,
into an energy independent potentials. This allow us to employ
the potential directly in configuration space  where 
various formalisms exist.  One such formalism, is the
IDEA method for few- and many-body systems  which,
apart from its simplicity, gives results which are 
in excellent agreement with those of other  methods
and therefore they can be safely employed in
our investigations on the in-medium few-body problem.\par

We demonstrate our approach by calculating
 the binding energies for the 2--, 3--, 4--, and 16--boson like systems
 at finite temperatures and densities.
The behavior of the adiabatic approximation potential surfaces,  
below and above the Mott density, is also investigated.
We found that the adiabatic potentials  exhibit similar 
features to those found in Ref. \cite{BS01}, namely, that 
the attractive well moves towards the interior region while 
at the same time is becomes narrower as the densities become larger.
The analytical properties of the corresponding Jost function
in the complex $k$-plane for  these potentials were investigated
in \cite{BS01}  where it was  shown that the bound state 
$E_b=-\hbar^2/2\mu\, b^2$ 
moves, downwards on the imaginary $k$-axis and eventually
become an anti-bound state instead of a resonance (as implied
by the appearance of a small repulsive  hump in the interaction 
region).  This reflects a vanishing imaginary part of the
respective two-body spectral function.  Hence, the two-body cluster
(even for $b<0$) retains the character of a quasi-particle, i.e. the
correlation does not decay. In a more general approach the
self-consistency requirement~\cite{Bozek:1999su} as well as three-body
collisions~\cite{Beyer:1997sf} lead to imaginary parts of the two-body
spectral function, and therefore to a finite width (life time) of the
two-particle cluster irrespective of $b$ smaller or larger than zero.\par

It is interesting to note that the change of the eigen-potential
for the 16-body  shows a lesser sensitivity
to the density changes and the binding energies are
still quite deep.  Further investigations with other 
bosonic systems and two-body
forces are needed to elucidate the behavior  and phase
transition of large clusters in the medium.

\vfill

%

\end{document}